\documentclass[twocolumn]{sig-alternate}
\usepackage[utf8]{inputenc}
\usepackage{amsmath,amsfonts,hyperref,bm,physics, amsthm}
\usepackage{amssymb}
\usepackage{xypic}
\usepackage{graphicx}
\usepackage{mathrsfs}
\usepackage{booktabs}
\usepackage{verbatim}
\usepackage{caption,afterpage}

\DeclareMathOperator*{\argmin}{arg\,min}

\usepackage[ruled,algo2e]{algorithm2e}
\usepackage{tikz}
    \usepackage{braket}

\theoremstyle{plain}

\theoremstyle{definition}

\theoremstyle{remark}

\renewcommand{\delta}{\varepsilon}

\providecommand{\RR}{\mathbb{R}}

\providecommand{\bf}{\mathbf{f}}

\providecommand{\bm}{\mathbf{m}}

\date{}

\title{A partial differential equations approach to defeating partisan gerrymandering}
\author{Matt Jacobs \thanks{MJ was supported by DARPA award FA8750-18-2-0066 and NSF grant DMS-1737770.} \and Olivia Walch}

\begin{document}
\maketitle

\abstract{We introduce a novel partial differential equations approach for addressing the problem of partisan gerrymandering. Our method is based on volume preserving curvature flow, a partial differential equation which we adapt to smooth voting district boundaries while preserving equal voting populations.  We show that every step of the flow minimizes a ``compactness energy'', allowing us to demonstrate that our method produces more ``compact'' and reasonable district maps.
We compute the flow using a variant of ``auction dynamics'' --- an efficient MBO type algorithm for computing volume preserving curvature flows.  This ``auction dynamics'' approach 
can be used to generate hundreds of reasonable maps in a matter of seconds without parallelization. The compactness energy provides a way of comparing proposed districtings of a given state. We demonstrate both the map generation and map comparison features of our approach for several different states.
}

\section{Introduction}

Partisan gerrymandering occurs when district lines are drawn to selectively dilute the voting power of minority groups and political opponents. The practice has a long history in the United States, with records of politically motivated district lines dating back to before the election of the First U.S. Congress in 1789 \cite{hunter2011}. In recent years, however, the problem has taken on much greater urgency \cite{whitford,lowv2018}. Sophisticated software and data mining techniques have allowed for the gerrymandering of voting districts with unprecedented efficacy.

A hallmark of a gerrymandered map is a collection of bizarrely drawn district boundaries, which serves to either concentrate or isolate voters of a particular political persuasion (``packing and cracking'').  As a result, many metrics for quantifying gerrymandering have focused on detecting unnatural shapes and boundaries. At least thirty distinct ways of scoring ``district compactness'' exist in the literature to-date, with many focusing on perimeter length and minimum bounding areas as key indicators of a district's compactness (\cite{reock1961,polsby1991,chambers2010,niemi1990}, among others). These scores have proven useful in making and bolstering legal arguments concerning gerrymandering, but leave many things to be desired \cite{barnes2018}. Moreover, the question of how to draw better districts is often left unaddressed, or can only be handled via computationally expensive random sampling methods \cite{fifield2015,Bangia2017}.

In this work, we treat the drawing of district lines as a partitioning problem.  This allows us to combat gerrymandering by drawing on techniques from the vast mathematical literature on partitioning. In recent years, partial differential equations (PDEs) have proven to be an invaluable tool for solving partitioning problems ranging from image segmentation \cite{Bresson2007, esedoglu2006} to abstract partitioning of graphs \cite{bertozzi2012, bresson2013, merkurjev2013,garciacardona2014}.  PDE-based approaches use a very simple idea: define an energy which captures the features of a desirable partition and then compute the gradient flow of the energy. The gradient flow follows the path of steepest descent for the energy. Thus, as the flow evolves in time the features of the partition improve.  In the context of gerrymandering, this approach provides a way to both score existing maps and generate new maps.

The heart of our method is a well-known geometric PDE: \emph{motion by mean curvature}. Motion by mean curvature or curvature flow acts on sets by evolving their boundaries.  Curvature flow arises as gradient descent for energies measuring the length of the boundary.  Thus, the PDE naturally acts to smooth and shorten boundaries between sets.   To produce valid district maps we must ensure that the districts maintain approximately equal voting populations throughout the evolution.  By interpreting district population as a volume constraint for the underlying set, we can maintain equal populations using \emph{volume preserving curvature flow}. 

Our approach to the districting problem is as follows. First, we define the districting problem over the structure of a weighted graph. The nodes of the graph are small community subunits defined by the U.S. Census Bureau. This allows us to partition a state into districts, while respecting local community structure. Next, we define a ``compactness energy'' which measures district boundaries and penalizes district sprawl. Finally, we compute the gradient flow of our energy using \emph{auction dynamics} --- a highly efficient algorithm introduced in \cite{jacobs2018} for computing multiphase volume preserving curvature flows.  
 
Our resulting algorithm is extremely efficient --- we can compute the entire trajectory of the flow in a matter of seconds even for large states. Since the flow is tied to an energy, we can naturally assign a score (which is provably minimized at each step) to every districting along the trajectory of the motion. Our approach allows us to impose very strict population constraints without difficulty, and the graph-based formulation of districting we employ allows us to enforce restrictions on the solution space without interrupting the flow. In what follows, we will describe our algorithm, discuss how it can be adapted for the problem of scoring and generating district maps, and provide examples of its output on several different states.

\section{Methods}

\subsection{Defining a districting}
We consider the district partitioning problem over the structure of a weighted graph $(\mathcal{V}, W)$.  The vertices $x\in\mathcal{V}$ of the graph are small community subunits, defined by the U.S. Census Bureau.  The geographical centroid of a subunit $x\in \mathcal{V}$ is denoted $c(x)$ which is a vector in $\RR^2$.  

 The weight matrix $W$ controls how strongly different subunits $x, y\in\mathcal{V}$ are connected to each other. We build the weights $W(x,y)$
according some distance relationship $d(x,y)$ between the centroids $c(x)$ and $c(y)$, such as the $L^2$ distance or the driving time distance.  If $x$ is one of the $k$ nearest neighbors of $y$, or $y$ is one of the $k$ nearest neighbors of $x$ (according to the chosen distance $d(x,y)$) we choose the Zelnick-Manor and Perona \cite{zelnikmanor2004} weights
\begin{equation} W(x,y)=\exp(-\frac{d(x,y)^2}{\sigma(x)\sigma(y)}) \end{equation}
otherwise we set $W(x,y)=0$.
Here $\sigma(x)$ is a local distance rescaling factor.  We define $\sigma(x)$ to be the distance between $x$ and its $\frac{k}{2}$th nearest neighbor.  By only connecting $k$ nearest neighbors, we ensure that $W$ is a sparse matrix.  Finally, we renormalize $W$ by taking $W\mapsto R^{-1/2}WR^{-1/2}$, where $R$ is the diagonal matrix of row sums of $W$, i.e. $R(x,x)=\sum_{y\in\mathcal{V}}W(x,y)$.

An $N$-district map is a partition of the vertices $x\in \mathcal{V}$ into $N$ pairwise disjoint districts $D_1, \ldots, D_N$.  To uphold the principle of ``one person, one vote,'' the populations $P_i$ of the districts must be approximately equal. We shall enforce this principle by placing a lower bound $P$ on the  population of each district.  A districting $D_1, \ldots, D_N$ is valid if $P_i\geq P$ for all districts $1\leq i\leq N$.  We shall let $S_N$ denote the set of all valid $N$-district partitions of $\mathcal{V}$. 

To determine the population $P_i$ of a district $D_i$ we must sum up the populations of each subunit $x\in D_i$.  The Census Bureau provides population data $p(x)$ for each $x\in\mathcal{V}$.  Thus,
$P_i=\sum_{x\in D_i} p(x).$ 
Note that the populations $p(x)$ may vary wildly across different subunits.  For example, urban census subunits have much larger populations than rural census subunits.  To avoid computational difficulties, we shall assume that each subunit contains at least one person.

In what follows, it will be convenient to represent a districting as a function $\bm{u}:\mathcal{V}\to  \RR^N$.  We can encode the partition by defining the entries $\bm{u}(x)=(u_1(x), \ldots, u_N(x))$ to be
\begin{equation*}\label{eq:u_partition}
u_i(x)=\begin{cases}
1&\textrm{if} \; x\in D_i\\
0 &\textrm{otherwise}.
\end{cases}
\end{equation*}

\subsection{The compactness energy}

At the heart of our approach is an energy functional measuring the ``compactness'' of a districting.  Our energy consists of two terms. The first approximately measures the boundary length of the districting, and the second penalizes district sprawl by measuring the distance between the centroids $c(x)$ and a central district cluster point $c_i$.  The relative importance of the two terms is balanced by a nonnegative parameter $\alpha$.  The energy is given by
\begin{equation}\label{eq:ms_hc}
\begin{split}
 E_{\alpha}(\bm{u},\bm{c})=\frac{1}{2}\sum_{i=1}^N\sum_{x,y \in \mathcal{V}} u_i(x)A(x,y)\big(1-u_i(y)\big) \\ +\alpha\sum_{i=1}^N \sum_{x\in\mathcal{V}} u_i(x)\lVert c_i -c(x) \rVert^2 
 \end{split}
 \end{equation}
 where $A=W^TW$. Note that the first term is equivalent to summing $\sum_{(x,y)\in \mathcal{E}_{\textup{cut}}} A(x,y)$
where $\mathcal{E}_{\textup{cut}}$ is the set of pairs of vertices $(x,y)$ placed in different districts. Thus, this term is exactly the graph cut with respect to the matrix $A$, which in turn approximates the length of the district boundaries.

 \subsection{MBO and auction dynamics }
 
 To compute the volume preserving gradient flow of (\ref{eq:ms_hc}) we turn to the auction dynamics algorithm introduced in \cite{jacobs2018}. Auction dynamics is a variant of the celebrated Merriman, Bence and Osher (MBO) algorithm \cite{merriman1992} for computing motion by mean curvature. 
 
 Auction dynamics computes the gradient flow of (\ref{eq:ms_hc}) using the variational framework developed by Esedo\={g}lu and Otto in \cite{esedoglu2015}.  Esedo\={g}lu and Otto showed that in the continuum setting, the MBO algorithm has an associated Lyapunov functional, the heat content energy.  From this viewpoint, each iteration of the MBO algorithm is equivalent to minimizing the linearization of the heat content at the current configuration.  Auction dynamics extends the scheme to volume preserving curvature flow by imposing a volume constraint on the minimization problem.
 
The first term in the compactness energy (\ref{eq:ms_hc}]) is the natural extension of the heat content energy to the graph setting.  Thus, we can obtain a ``graph auction dynamics scheme'' by minimizing linearizations of (\ref{eq:ms_hc}) over the set of admissible partitions $S_N$.   In other words, we compute a gradient flow of our energy by iterating:

\begin{equation}\label{eq:lin_min} \bm{u}^{n+1}= \argmin_{\bm{u}\in S_N} \, (\nabla E_{\alpha}(\bm{u}^n,\bm{c}), \bm{u}-\bm{u}^n). \end{equation}

Note that $A=W^TW$ is a positive definite matrix, thus $E_{\alpha}$ is a concave function of $\bm{u}$.  The concavity guarantees that $E_{\alpha}$ lies below the tangent plane at any point $\bm{v}$.  Therefore,
\begin{equation}\label{eq:ghc_lin_trick} E_{\alpha}(\bm{u},\bm{c})\leq E_{\alpha}(\bm{v},\bm{c}) + (\nabla E_{\alpha}(\bm{v},\bm{c}), \bm{u}-\bm{v})\end{equation}
for any points $\bm{u}$ and $\bm{v}$.   This inequality guarantees that $E_{\alpha}$ is non-increasing at every step of the scheme.

  Equation (\ref{eq:lin_min}) is a non-trivial combinatorial optimization problem.  Although the objective function is linear, the set of admissible partitions, $S_N$, is a non-convex set.  To overcome this difficulty, we can instead pose the minimization over the convex hull of $S_N$.   This relaxation transforms problem (\ref{eq:lin_min}) into a linear programming problem 
\begin{equation}\label{eq:lin_program}
\begin{split}
\bm{u}^{n+1}=\argmin_{\bm{u}\geq 0} \,(\nabla E_{\alpha}(\bm{u}^n,\bm{c}), \bm{u}-\bm{u}^n) \; \\ 
\textrm{s.t.} \, \sum_{i=1}^N u_i(x)=1,  \, \sum_{x\in\mathcal{V}} u_i(x)p(x)\geq P. 
\end{split}
\end{equation}

 When the populations $p(x)$ are uniform, the solution to the original combinatorial optimization problem (\ref{eq:lin_min}) coincides with the solution to (\ref{eq:lin_program}).  In our case, $p(x)$ is not uniform; thus, the solution to (\ref{eq:lin_program}) may not be a partition, i.e. $\bm{u}(x)$ may have fractional entries for some $x\in \mathcal{V}$.  If $\bm{u}(x)$ has fractional entries, then the vertex $x$ is split between two (or more) different districts --- an undesirable outcome.

In practice, replacing problem (\ref{eq:lin_min}) with the relaxation (\ref{eq:lin_program}) rarely leads to vertex splitting.  This is unsurprising given that both terms in the compactness energy (\ref{eq:ms_hc}) increase if some $\bm{u}(x)$ has fractional entries. As a result, whenever an iterate $\bm{u}^n$ develops fractional entries, the gradient $\nabla E_{\alpha}(\bm{u}^n,\bm{c})$ will point in a direction which encourages $\bm{u}^{n+1}$ to have binary entries.  Thus, if a vertex $x$ is split in the course of the flow, it will be quickly corrected in the next few iterations.    

Auction dynamics solves subproblem (\ref{eq:lin_program}) using the auction algorithm introduced by Bertsekas in \cite{bertsekas1979} (hence the name ``auction dynamics'').  See \cite{jacobs2018} for an exhaustive discussion of auction algorithms in the context of multiphase volume constrained curvature flow. 

After each $\bm{u}$ update step, we update the district centroids $c_i$ to minimize $E_{\alpha}$. $E_{\alpha}$ is quadratic in $\bm{c}$, thus the update
\begin{equation} \label{eq:centroid_update} \bm{c}^{n+1}=\argmin_{\bm{c}\in\RR^{N \times 2}} E_{\alpha}(\bm{u}^{n+1}, \bm{c})\end{equation} 
is easily calculated. 

We are now ready to give our auction dynamics algorithm, Algorithm \ref{alg:main_alg}, for redistricting.  The algorithm alternates computing gradient flow of energy (\ref{eq:ms_hc}) with respect to the partition $\bm{u}$ and updating the district centroid $\bm{c}$ (Equations (\ref{eq:lin_program}) and (\ref{eq:centroid_update}) respectively).

\subsection{Temperature}
To prevent the algorithm from computing the same trajectory every time, we can introduce randomness into the entries of the gradient vector $\psi_i^{n+1}(x)$.  We interpret this randomness as fluctuations due to temperature.  For some temperature level $T$ we add independent identical Gaussian random variables with distribution $\mathcal{N}(0,T)$ to $\psi_i^{n+1}(x)$ for every $i$ and $x$ before running the auction step.  Note that adding temperature can allow the trajectory to escape local minima and find potentially more compact maps. 

Since we want to eventually find low energy maps, it makes sense to anneal the temperature during the flow.  Once the temperature is functionally zero, the energy is certain to be decreased at each step. When we include temperature in our simulations, we always anneal it at a fixed rate; e.g. decreasing it to 95\% of its value at the prior step.

\begin{algorithm2e}
Compute the gradient:
\begin{equation} \psi^{n+1}_i(x)=\alpha \lVert c^n_i-c(x)\rVert^2-\sum_{y\in \mathcal{V}} W(x,y)u^{n}_i(y) \end{equation}
Solve the linear program with an auction:
\begin{equation}
\begin{split}
\bm{u}^{n+1}=\argmin_{\bm{u}\geq 0} \sum_{i=1}^N \sum_{x\in \mathcal{V}} u_i(x)\psi^{n+1}_i(x) \; \\ 
\textrm{s.t.} \, \sum_{i=1}^N u_i(x)=1,  \, \sum_{x\in\mathcal{V}} u_i(x)p(x)\geq P 
\end{split}
\end{equation}
Update the district centroids:

\begin{equation} 
c_i^{n+1}=\frac{\sum_{x\in \mathcal{V}}u^{n+1}_i(x) c(x)}{\sum_{x\in\mathcal{V}}  u^{n+1}_i(x)}
\end{equation}
\caption{Auction dynamics}\label{alg:main_alg}
\end{algorithm2e}

\subsection{Data}
All census tract shapefiles were obtained from the Census website, as were the Congressional District shapefiles. Population data for each tract was taken from the American Community Survey data available via the American FactFinder website (\url{https://factfinder.census.gov/}). Code and data for all figures is available on Github (\url{http://github.com/ojwalch/district_mbo}).

\section{Results}

\begin{figure*}
  \includegraphics[width=\linewidth]{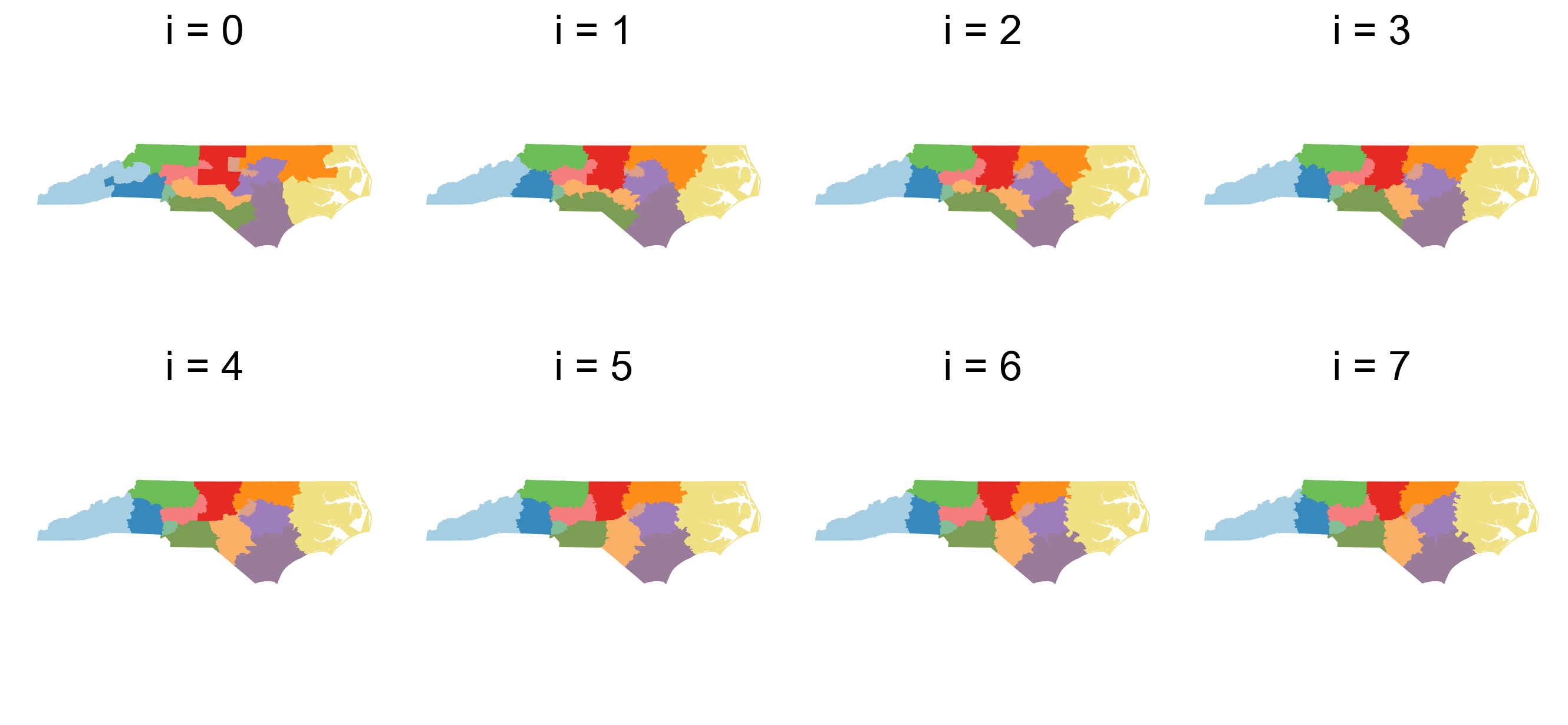}
  \caption{Sample flow in North Carolina. The initial districting (upper left) is updated in each of the subsequent maps (read left-to-right) according to motion by mean curvature. In this simulation, the number of nearest neighbors is $k = 150$, the centroid distance coefficient $\alpha = 1$, and there is no temperature. }
\label{fig:flow}
\end{figure*}

The modified auction dynamics algorithm is able to rapidly produce large numbers of compact, contiguous districtings.
Even for large states such as Ohio or Pennsylvania, we are able to compute the flow in a matter of seconds.  The algorithm is robust to initial conditions, producing reasonable districtings even when starting from a random partition. 

\begin{figure}
  \includegraphics[width=\linewidth]{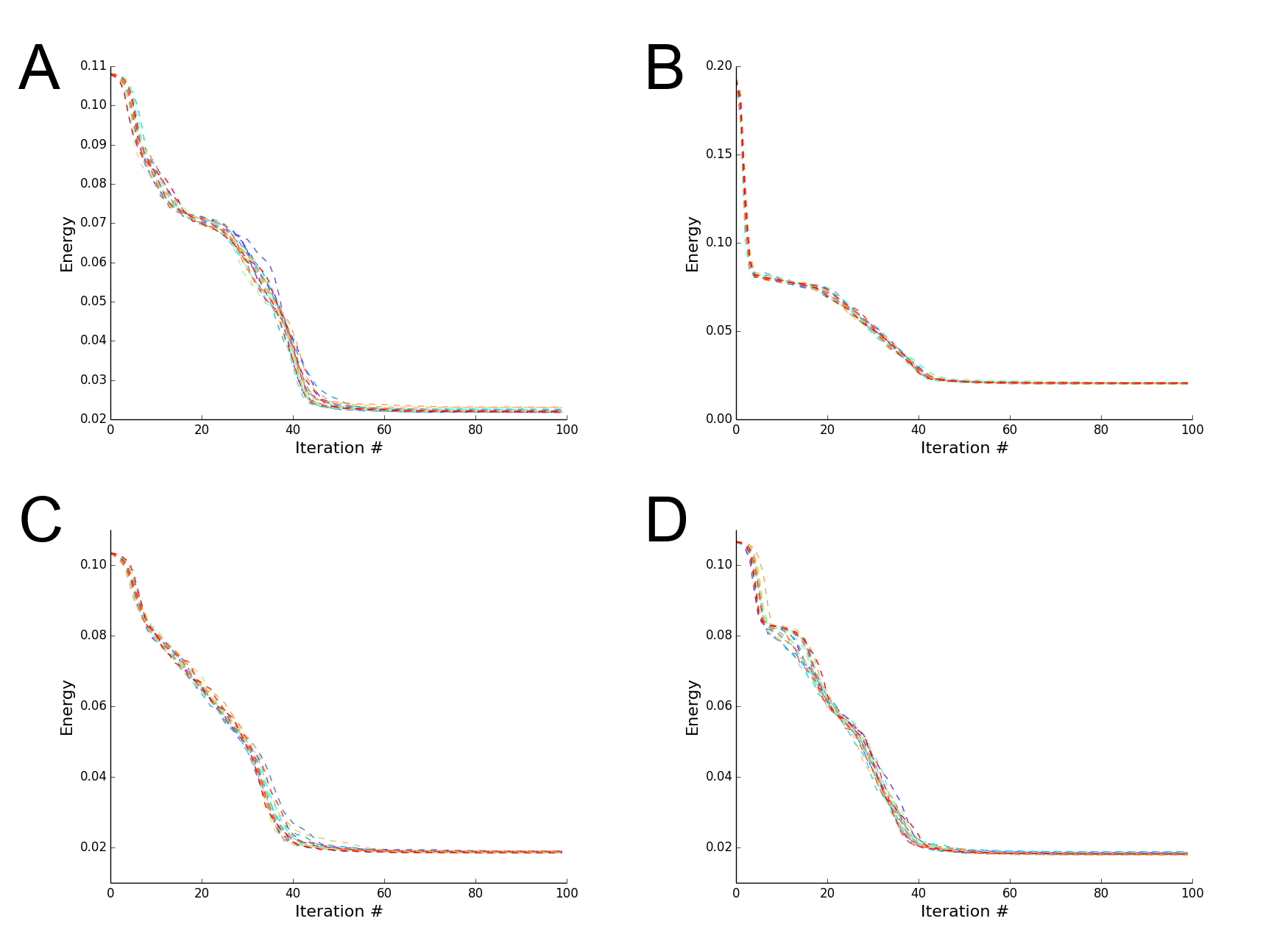}
  \caption{Change in energy across 100 iterations for 20 runs of the algorithm on several states, with blocks initially assigned to districts at random. (A) Michigan, (B) Pennsylvania, (C) North Carolina, (D) Virginia. Temperature is set to 0.1 at the beginning for all trials and anneals to 95\% its previous value with every iteration.}
\label{fig:energy}
\end{figure}

Energy decreases with each iteration of the algorithm, except in cases when temperature is included. As temperature is always annealed in our approach, the energy will decrease monotonically once temperature is effectively zero. A sample flow from North Carolina is shown in Figure~\ref{fig:flow}. Energy trajectories for several states are plotted in Figure \ref{fig:energy}. 
For each state, the trajectories are remarkably similar, despite different initial conditions and random fluctuations due to temperature. Thus, we see that the algorithm very reliably finds low energy configurations.  The final maps, while similar in energy, can look very dissimilar.

\begin{figure}
\includegraphics[width=\linewidth]{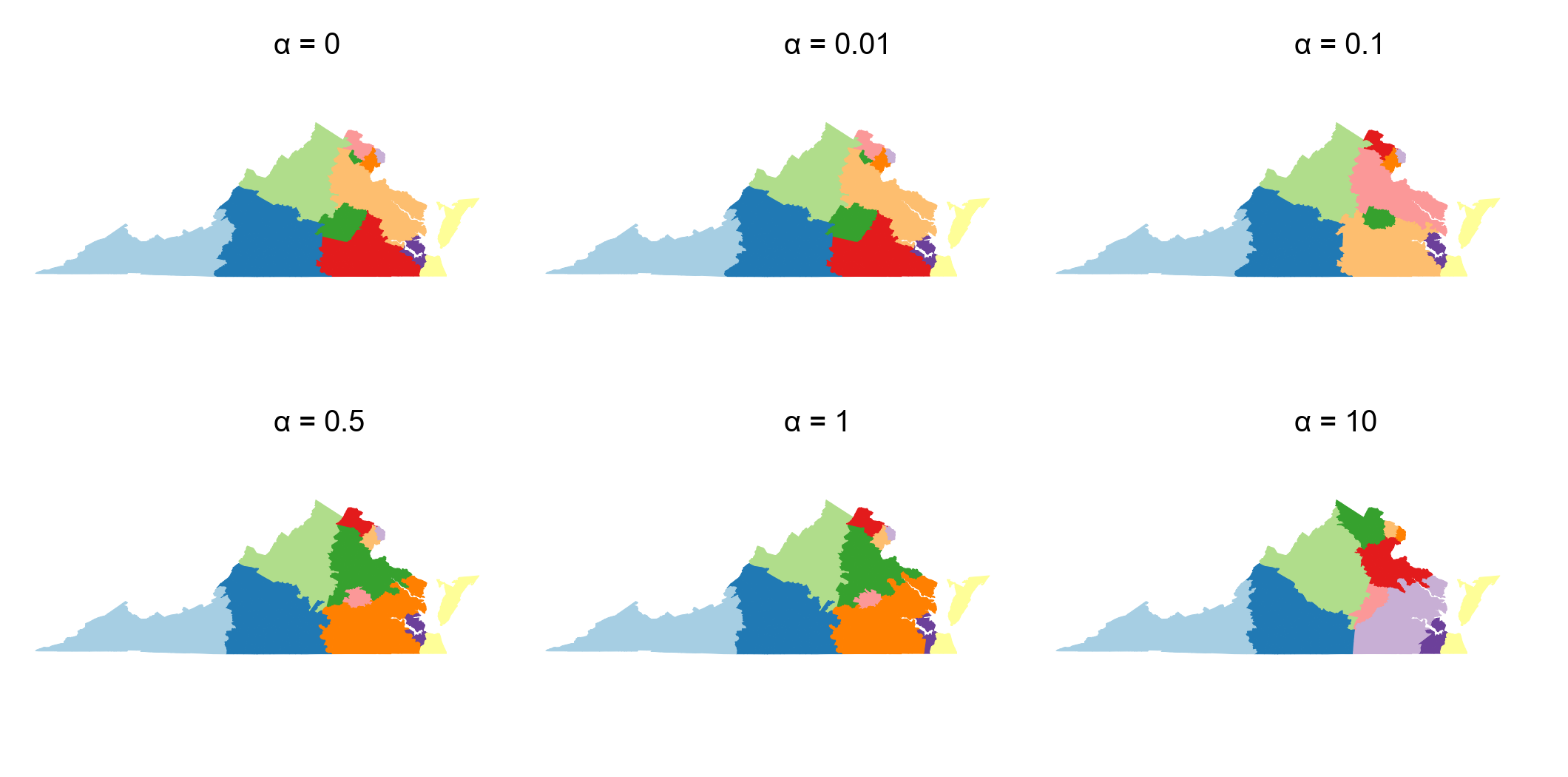}
\caption{Effects of varying $\alpha$ on six separate runs of our algorithm. Output maps for Virginia are shown with increasing centroid distance parameter. Each map is initialized from the 2016 districting and run without temperature. The only parameter that varies across maps is the centroid distance coefficient $\alpha$ (values shown in Figure) that penalizes dispersed districts, with lower $\alpha$ linked to a higher occurrence of non-contiguous districts.}
  \label{fig:ms}
\end{figure}

\begin{figure}
  \includegraphics[width=\linewidth]{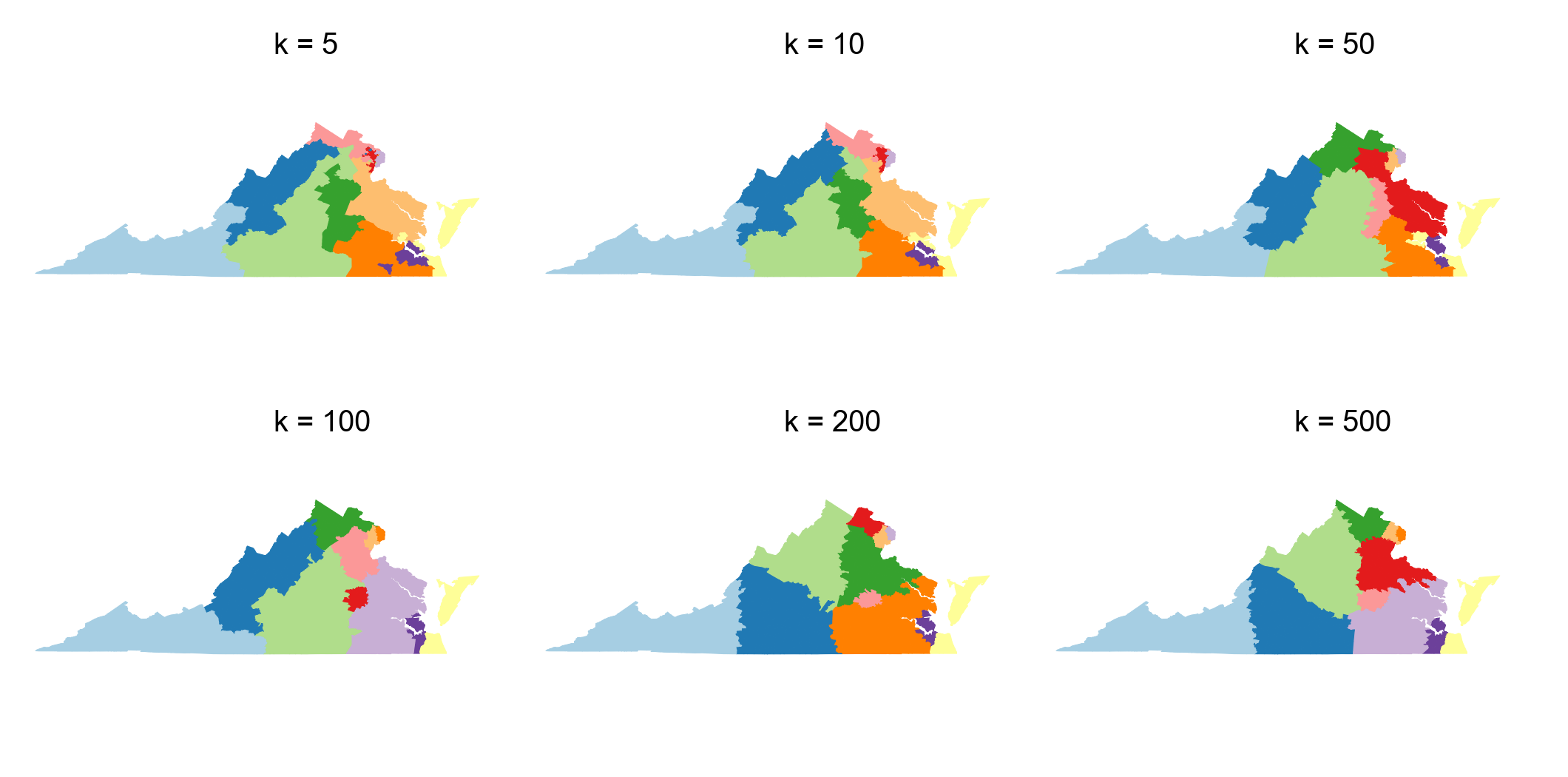}
  \caption{Choice of $k$ on mapping. The number of nearest neighbors is varied from 5 to 500. Each simulation is run without temperature, starting from the 2016 districting. If $k$ is too small, the algorithm will not produce a contiguous mapping. Above 100, the choice of $k$ appears not to play a significant role in the outcome. }
  \label{fig:knn}
\end{figure}

\subsection{Choice of parameters}
In the absence of temperature, the key parameters to choose for the algorithm are $\alpha$, the centroid distance coefficient, and $k$, the number of nearest neighbors.   

The parameter $\alpha$ controls the relative importance of the graph cut term and the centroid penalty term.  Without a centroid penalty term, curvature flow can sometimes separate a set into two (or more) connected components.  Separation leads to a high centroid penalty, thus separation can be prevented by choosing $\alpha$ sufficiently large. 

Figure \ref{fig:ms} shows the output of the algorithm with several different values of $\alpha$. For small values of $\alpha$ some of the districts become disconnected. As $\alpha$ is increased, the separations disappear and all of the computed districts are contiguous.
For values greater than $\alpha = 2$, the algorithm does not seem to yield markedly different qualitative results for different values of $\alpha$ (see Figure \ref{fig:ms}). 

It is important to choose $k$ sufficiently large. The weight matrix must have enough entries so that each district can ``see'' the neighboring vertices it wants to capture. If there are too few nearest neighbors, the algorithm will be slow or completely stationary.    For values of $k \geq 100$ the algorithm reliably converges (Figure \ref{fig:knn}). 

An additional parameter to consider is the lower bound for the populations of each district, $P$. In our simulations, we set $P =  \,  \frac{0.999}{N}\sum_{x\in\mathcal{V}} p(x)$ i.e. each district must have at least $99.9\%$ of the total population divided by the number of districts. Allowing a small amount of flexibility in the population constraint ensures that the algorithm can find valid configurations at each step.

\begin{figure}
  \includegraphics[width=\linewidth]{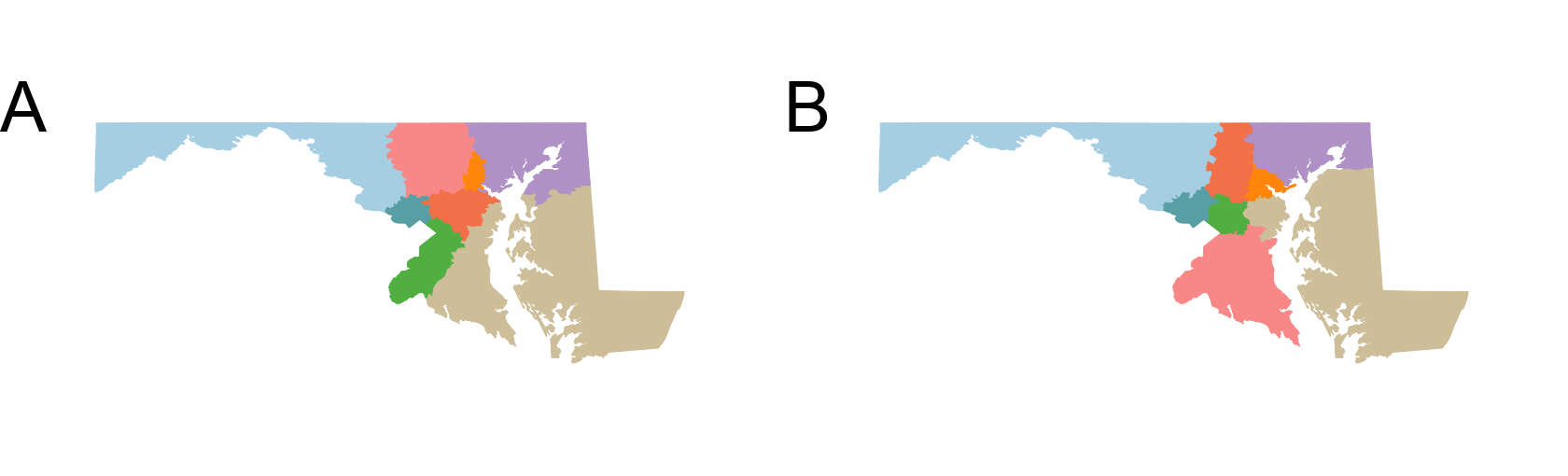}
  \caption{Choice of weight source affects final map output. On the left, the result from applying our algorithm to Maryland using the $L^2$ distance to construct weights. 
  On the right, the weights are determined by driving time between two block centroids, as calculated using the Open Source Routing Machine (\url{http://project-osrm.org}). A bridge connects the tan-colored district's noncontiguous components in Map B. In general, human-centric measures of distance produce more natural results.}
  \label{fig:driving_distance}
\end{figure}

\subsection{Choice of distance measure}
In most of our simulations, we choose the usual $L^2$ metric to define the graph weights, but this is not required. Alternative distances can be used that more accurately reflect the human geography of the map, yielding more reasonable and useful mappings.

Maryland is an illuminating case study for alternative measures of distance, as standard $L^2$ distances make it so that tracts separated by the Chesapeake Bay are connected (Figure \ref{fig:driving_distance}A). This is resolved by using driving time as the distance measure.  With the driving time distance the only district spanning the water (tan-colored) does so because of the existence of a nearby bridge (Figure \ref{fig:driving_distance}B). 

The distance between different units $x$ and $y$ can also be artificially set to zero. Setting $d(x,y)=0$ encourages placing $x$ and $y$ in the same district. This could be desirable for Voting Rights Act compliance or to preserve historical communities.

\subsection{Comparing the energy of multiple maps}
Our energy (\ref{eq:ms_hc}) can be used as a compactness score for a given districting. Auction dynamics computes stationary points of the energy, which can serve as low-energy benchmarks. The compactness energy is non-convex (an inherent feature of any nontrivial partitioning problem), thus auction dynamics will converge to different stationary points depending on the initial configuration and random fluctuations due to temperature.  Hence, a histogram of many low-energy districtings provides a more robust benchmark for the compactness of a given map. 

\begin{figure}
  \includegraphics[width=\linewidth]{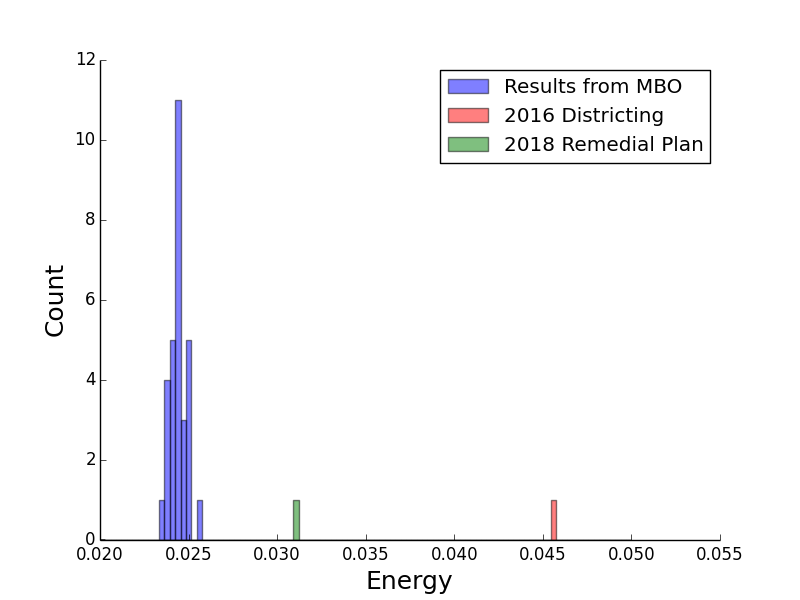}
  \caption{Comparing the energy of proposed districtings to the final energies resulting from our algorithm. The energy of Pennsylvania's 2016 Congressional Districting Plan (red) is compared to the 2018 Remedial PA Mapping (green), and the final energies of 30 separate runs of our algorithm starting from random initial districtings in Pennsylvania (blue). Across all trials the parameters used to compute the energy (e.g. number of neighbors $k$ and centroid distance coefficient $\alpha$) are the same.}
  \label{fig:energy_hist}
\end{figure}

In Figure \ref{fig:energy_hist}, the energies of the 2016 Pennsylvania mapping and the 2018 Remedial PA plan are compared to the output from 30 trials of our algorithm with random initialization (i.e. each vertex assigned to a random district). The Remedial plan (green) is significantly closer to the locus of low-energy solutions than the 2016 mapping (red), and this holds for all reasonable values of $k$ and $\alpha$.

\begin{figure}
  \includegraphics[width=\linewidth]{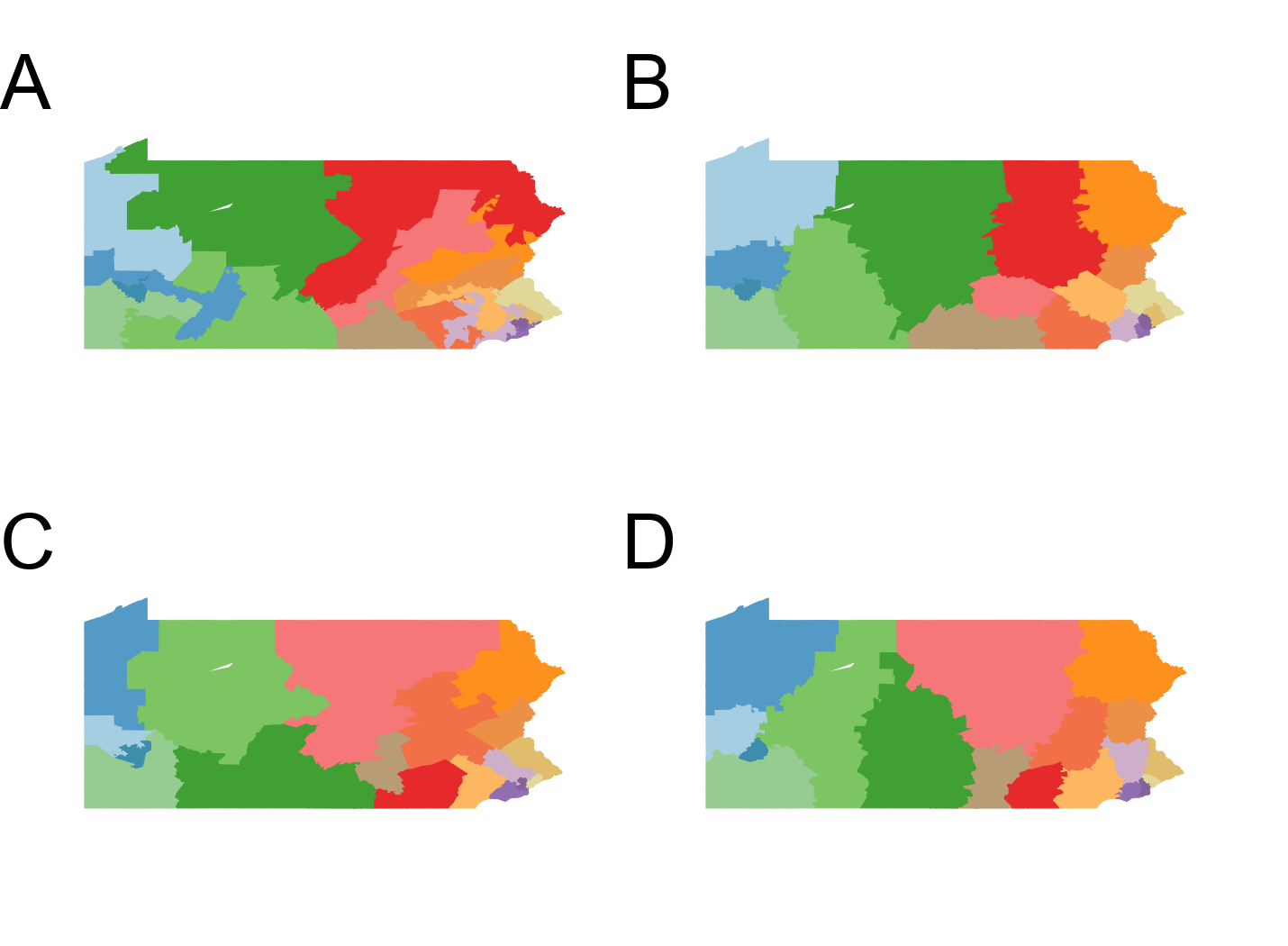}
  \caption{Before and after results of running the algorithm on Pennsylvania districtings. A) Pennsylvania's 2016 district mapping, B) the results of running the auction dynamics algorithm initializing from (A), C) the 2018 Remedial mapping, D) the results of running the auction dynamics algorithm initialized from (C). The 2016 mapping changes much more dramatically than the 2018 mapping.}
  \label{fig:pa}
\end{figure}

For insight into this result, one can look at the difference between the 
auction dynamics output when initialized with the 2016 map versus the 2018 remedial map (Figure \ref{fig:pa}). The remedial mapping is much closer to the final solution and changes much less dramatically over the course of the flow than the 2016 mapping.

\section{Conclusion}

In this paper, we adapt a partial differential equations technique for partitioning problems to address the specific conditions (equal populations, contiguity) required by districting. We introduce a compactness score, the compactness energy, and create new district maps by following the volume preserving gradient flow of the energy. This approach allows us to both score the compactness of a districting and to generate large numbers of compact and contiguous mappings that can be used to provide context to any proposed districting.
Our method is readily adaptable to real-world constraints: community units that are legally mandated to be in the same district can be handled by artificially decreasing the distance between them, and the weights can be chosen to capture more ``human'' metrics such as travel time distance.
The code runs quickly, does not require parallelization, and is open source.

Ultimately, the goal of districting is to produce a map which fairly represents the will of the populace.  The proposed algorithm is not a means of arriving at one ``optimal'' districting; rather, it is a tool that can be used to evaluate proposed districtings in the context of the constraints imposed by geography and population distribution.  If the current districting does a worse job representing the people than maps along the trajectory of the flow, it should be considered highly suspect and very likely gerrymandered. Future investigations will clarify the role for this technique in combating real-world partisan gerrymandering. 

\section{Acknowledgments}
This material is based on research sponsored by the Air Force Research Laboratory and DARPA under agreement number FA8750-18-2-0066.
The U.S. Government is authorized to reproduce and distribute reprints for Governmental purposes notwithstanding any copyright notation thereon.

The views and conclusions contained herein are those of the authors and should not be interpreted as necessarily representing the official policies or endorsements, either expressed or implied, of the Air Force Research Laboratory and DARPA or the U.S. Government.

\end{document}